\documentclass[conference]{IEEEtran}
\ifCLASSINFOpdf
\else
\fi
\usepackage{cite}
\usepackage{url}
\usepackage{graphicx}
\usepackage{subfigure}
\usepackage{tikz-cd}
\usepackage{tikz}
\usepackage{booktabs}
\usepackage{epstopdf}
\usepackage{bm}
\usepackage{amsmath}
\usepackage{amssymb}
\usepackage[numbers]{natbib}
\usepackage{color}
\usepackage{multirow}
\usepackage{tabularx}

\usetikzlibrary{positioning}
\usetikzlibrary{shapes,snakes}
\usepackage{algorithm}
\usepackage{algorithmic}

\begin{document}
\newcolumntype{L}[1]{>{\raggedright\arraybackslash}p{#1}}
\newcolumntype{C}[1]{>{\centering\arraybackslash}p{#1}}
\newcolumntype{R}[1]{>{\raggedleft\arraybackslash}p{#1}}
%
\title{SINE: Scalable Incomplete Network Embedding}




\author{\IEEEauthorblockN{Daokun Zhang\IEEEauthorrefmark{1},
Jie Yin\IEEEauthorrefmark{2},
Xingquan Zhu\IEEEauthorrefmark{3}, and
Chengqi Zhang\IEEEauthorrefmark{1}}
\IEEEauthorblockA{\IEEEauthorrefmark{1}Centre for Artificial Intelligence, FEIT, University of Technology Sydney, Australia\\
\fontsize{9}{9}\selectfont\ttfamily\upshape
Email: Daokun.Zhang@student.uts.edu.au; Chengqi.Zhang@uts.edu.au}
\IEEEauthorblockA{\IEEEauthorrefmark{2}Discipline of Business Analytics, The University of Sydney, Australia\\
\fontsize{9}{9}\selectfont\ttfamily\upshape
Email: Jie.Yin@sydney.edu.au}
\IEEEauthorblockA{\IEEEauthorrefmark{3}Department of Computer \& Electrical Engineering and Computer Science, Florida Atlantic University, USA\\
\fontsize{9}{9}\selectfont\ttfamily\upshape
Email: xqzhu@cse.fau.edu}
}


\maketitle

\begin{abstract}
	Attributed network embedding aims to learn low-dimensional vector representations for nodes in a network, where each node contains rich attributes/features describing node content. Because network topology structure and node attributes often exhibit high correlation, incorporating node attribute proximity into network embedding is beneficial for learning good vector representations. In reality, large-scale networks often have incomplete/missing node content or linkages, yet existing attributed network embedding algorithms all operate under the assumption that networks are complete. Thus, their performance is vulnerable to missing data and suffers from poor scalability.
	
	In this paper, we propose a Scalable Incomplete Network Embedding (SINE) algorithm for learning node representations from incomplete graphs. SINE formulates a probabilistic learning framework that separately models pairs of node-context and node-attribute relationships. Different from existing attributed network embedding algorithms, SINE provides greater flexibility to make the best of useful information and mitigate negative effects of missing information on representation learning. A stochastic gradient descent based online algorithm is derived to learn node representations, allowing SINE to scale up to large-scale networks with high learning efficiency. We evaluate the effectiveness and efficiency of SINE through extensive experiments on real-world networks. Experimental results confirm that SINE outperforms state-of-the-art baselines in various tasks, including node classification, node clustering, and link prediction, under settings with missing links and node attributes. SINE is also shown to be scalable and efficient on large-scale networks with millions of nodes/edges and high-dimensional node features. The source code of this paper is available at https://github.com/daokunzhang/SINE.

\end{abstract}


%
\IEEEpeerreviewmaketitle

\section{Introduction}

Network embedding, also known as network representation learning, aims to embed each node of a network into a low-dimensional vector space, by preserving network structure and other side information. As a result, network analytical tasks can be easily conducted by applying machine learning techniques to the new vector space. Due to the increasing popularity of networked applications, such as social networks, real-world networks are often of large scale, containing a large number of nodes, links, and high-dimensional content features. These challenges have motivated the development of many network embedding solutions in the field.



Two main streams of network embedding algorithms include (1) \textit{structure preserving network embedding methods}, e.g., DeepWalk~\cite{perozzi2014deepwalk}, LINE~\cite{tang2015line}, node2vec~\cite{grover2016node2vec}, that preserve only network structure, and (2) \textit{attributed network embedding methods}, e.g., TADW~\cite{yang2015network}, HSCA~\cite{zhang2016homophily}, MVC-DNE~\cite{yang2017properties}, that augment network structure with node attributes. Because structure preserving network embedding methods only utilize network topology structure, they have shown to be inferior to attributed network embedding methods which combine both node content and structure information for embedding learning. Meanwhile, while existing attributed network embedding methods can leverage node content, they often assume the input networks are complete and cannot handle missing data. In addition, they suffer from poor scalability due to high computational cost. In summary, two major drawbacks of the existing attributed network embedding methods include:

\noindent\textbf{Vulnerable to missing data:} Real-world networks are often incomplete with missing edges and/or missing node content features~\cite{kossinets2006effects}, due to various reasons. First, privacy or legal restrictions make sensitive information on node attributes or part of connections among nodes inaccessible. Second, networks have too large size, making it prohibitively expensive or even impossible to directly acquire complete networks. Instead, a common practice is to obtain a smaller sample of large networks for analysis. The sampled network inevitably contains lots of missing nodes and links. Third, networks are dynamic in nature, and thus newly joined nodes often have very few links or content features. All these aspects result in noisy and incomplete networks. Although research has shown that jointly exploiting network structure and node attributes can enhance the embedding performance, existing attributed network embedding algorithms require node attributes to be all complete. When nodes in networks have no attributes or important dimensions of attributes are unobservable, existing methods are vulnerable to such missing data.

\noindent\textbf{Poor scalability:} Most attributed network embedding algorithms rely on matrix factorization (e.g., TADW~\cite{yang2015network}, HSCA~\cite{zhang2016homophily}) or deep neural networks (e.g., MVC-DNE~\cite{yang2017properties}) to fuse the information on network structure and node attributes towards learning a better joint representation. The matrix factorization or deep neural networks require high computational cost, where the time complexity is at least quadratic to the number of nodes, preventing them from scaling to large-scale networks with a large number of nodes and high-dimensional node features. Thus, there is a strong demand for developing efficient and effective network embedding algorithms.

The above observations motivate our research to find a better network representation that is not only robust to missing data in networks, but also scalable to large-scale networks.

In this paper, we propose a new attributed network embedding algorithm, called SINE, that provides a probabilistic formulation that 1) models pairs of node and context relationships to capture broader structural dependency in random walks; 2) models pairs of node and attribute relationships to make the best of available node content information. Different from existing attributed network embedding algorithms, SINE provides a flexible way to leverage useful information and diminish the negative impacts on the learned embedding representation, caused by the existence of missing data on network structure and/or node attributes. We derive an online optimization strategy based on stochastic gradient descent, which enables SINE with high learning efficiency and the ability to scale up to large networks. We evaluate the efficacy and efficiency of SINE on real-world networks in various tasks, including node classification, clustering, and link prediction. As compared with the state-of-the-art baselines, SINE is demonstrated to be robust to missing links and node attributes, but also scalable to large-scale networks.

The main contribution of this paper is threefold:
\begin{itemize}
	\item We advance the existing attributed network embedding learning to a realistic missing data setting, allowing embedding leaning to be highly efficient and accurate for real-world networks.
	\item We propose a scalable and efficient algorithm to combine network structure and node attributes to learn a joint embedding representation, thereby diminishing negative impacts of missing information.
	\item We evaluate the effectiveness of the proposed method under different missing data settings, showing its superior performance to the state-of-the-art baselines.
\end{itemize}

The remainder of this paper is organized as follows. Section~\ref{sec:related work} reviews previous work related to network embedding and incomplete network analysis. The problem definition and preliminaries directly related to our formulation are given in Section~\ref{sec:problem and preliminaries}, followed by the proposed algorithm described in Section~\ref{sec-methodology}. Experiments and discussions are presented in Section~\ref{sec-experiments}, and we conclude the paper in Section~\ref{sec-conclusion}.

\section{Related Work}
\label{sec:related work}

\subsection{Network Embedding}

Existing research on unsupervised network embedding can be divided into two categories~\cite{zhang2018network}: \textit{structure preserving network embedding} algorithms that only leverage network structure to learn node embeddings, and \textit{attributed network embedding} algorithms that couple network structure with node attributes for more effective network embedding.

\subsubsection{Structure Preserving Network Embedding} Inspired by Skip-Gram~\cite{mikolov2013distributed}, DeepWalk~\cite{perozzi2014deepwalk} learns node representations by preserving the similarity between nodes sharing similar contexts. Random walks are adopted by DeepWalk to obtain node contexts. node2vec~\cite{grover2016node2vec} (a variant of DeepWalk) exploits biased random walks to capture more flexible structural contexts. LINE~\cite{tang2015line} learns node embeddings by directly modeling the first-order proximity (the proximity between connect nodes) and the second-order proximity (the proximity between nodes sharing direct neighbors). GraRep \cite{cao2015grarep} steps further to consider high-order proximities by modeling the relations between nodes and their $k$-step neighbors. M-NMF~\cite{wang2017community} complements the local structural proximity with community structure to learn community preserving node representations. Deep learning techniques are also adopted to learn deep, non-linear node representations. DNGR~\cite{cao2016deep} employs stacked denoising autoencoder \cite{vincent2010stacked} to learn deep low-dimensional node embeddings. SDNE~\cite{wang2016structural} designs a semi-supervised autoencoder to learn node representations that preserve the first-order and second-order proximity.

These algorithms leverage only network structure to learn node representations, while ignoring node content information. In situations where the links are sparse, these algorithms fail to produce satisfactory results.

\subsubsection{Attributed Network Embedding}

TADW~\cite{yang2015network} incorporates node text features into network embedding through inductive matrix factorization~\cite{natarajan2014inductive}. HSCA~\cite{zhang2016homophily} enhances TADW through enforcing the first-order proximity in the embedding space. pRBM~\cite{wang2016paired} constructs node representations from node attributes with Restricted Boltzmann Machine~\cite{hinton2006reducing}, and simultaneously preserves the similarity between connected nodes. To deal with social networks with noisy user profile features, UPP-SNE~\cite{zhang2017user} learns node representations by performing a structure-aware non-linear mapping on user profile features. CANE~\cite{tu2017cane} learns context-aware node embeddings from node attributes via the mutual attention mechanism. MVC-DNE~\cite{yang2017properties} adopts deep multi-view learning to learn node representations that encode network structure and node content features. GraphSAGE~\cite{hamilton2017inductive} infers node representations inductively from node content features through neighborhood feature aggregation. AANE~\cite{huang2017accelerated} learns node representations by finding a low-dimensional content feature subspace, where the distance between connected nodes is penalized.

The above attributed network embedding methods assume that networks are complete. When missing data is present in the networks, especially on node attributes, their performance deteriorates dramatically, because they lack the ability to handle missing data. In addition, because these methods are mostly based on matrix factorization and deep learning techniques, most of them also suffer from poor scalability. In contrast, our work proposes an effective and efficient way to learn node embeddings robust to missing data.

\subsection{Incomplete Graph Mining}

Missing data is very common in networks, but few research has investigated incomplete network mining. It is well recognized that, if missing data is simply ignored, network analysis results will be severely skewed. Therefore, research efforts have been put on missing data imputation before network analysis is performed. \cite{kim2011network} studies the network completion problem, where the focus is to learn a probabilistic model that fits the observed part of a network, and then uses the model to infer missing nodes and links of the network. More specifically, \cite{sundareisan2015hidden} addresses the problem of recovering the missing infections and the source nodes of an epidemic from sampled snapshots of large graphs. The notion of graph identification is introduced in \cite{jr2009identifying}, which aims to infer a cleaned output network from a noisy, incomplete input graph. \cite{smith2018network} studies the effects of non-random missing data on common network measures such as centrality, homophily, topology and centralization. However, very little attention has been put on investigating network embedding and developing robust algorithms for large-scale incomplete networks.

\section{Problem Definition and Preliminaries}
\label{sec:problem and preliminaries}

\subsection{Problem Definition}

Assume an incomplete network $G=(\mathcal{V},\mathcal{E},\mathcal{A},X, \Omega)$ is given, where  $\mathcal{V}$ is the set of nodes, $\mathcal{E}$ is the set of observed edges, and $\mathcal{A}$ is the set of node attributes. $X\in\mathbb{R}^{|\mathcal{V}|\times|\mathcal{A}|}$ is node feature matrix, with each observed element $X_{ij}\geqslant 0$ indicates the occurrence times/weights of attribute $a_{j}\in\mathcal{A}$ at node $v_{i}\in\mathcal{V}$, and $\Omega\in{\{0,1\}}^{|\mathcal{V}|\times|\mathcal{A}|}$ indicates whether the node attribute value $X_{ij}$ is observed, with $\Omega_{ij}=1$ for observed $X_{ij}$ and $\Omega_{ij}=0$ for unobserved $X_{ij}$. For networks with attributes taking continuous values, discretization can be applied to convert the continuous attributes to discrete ones.

The objective of incomplete network embedding is to leverage incomplete information in $\mathcal{E}$ and $X$ to learn a mapping function $\mathrm{\Phi}: v_{i}\in\mathcal{V}\mapsto \mathbb{R}^{|V|\times d}$. The learned node representations $\mathrm{\Phi}(v_{i})$ are expected to be (1) low-dimensional with $d\ll|\mathcal{V}|$, and (2) informative for the downstream tasks, such as node classification, node clustering and link prediction, etc.

\subsection{DeepWalk}

The Skip-Gram model~\cite{mikolov2013distributed} learns word representations by capturing the semantic similarity between words sharing similar contexts. DeepWalk generalizes the idea of Skip-Gram from word representation learning to network embedding, by using random walks to collect node contexts. Given a truncated random walk with length $L$, $\{v_{r_{1}}, v_{r_{2}},\cdots,v_{r_{i}},\cdots v_{r_{L}}\}$, DeepWalk learns representation $\mathrm{\Phi}(v_{r_{i}})$ for node $v_{r_{i}}$ by using it to predict its context nodes, which is achieved by solving the following optimization problem,
\begin{equation}
\min_{\mathrm{\Phi}}-\log\mathrm{P}(\{v_{r_{i-t}}, \cdots, v_{r_{i+t}}\}\setminus v_{r_{i}}|v_{r_{i}}),
\end{equation}where $\{v_{r_{i-t}}, \cdots, v_{r_{i+t}}\}\setminus v_{r_{i}}$ are the context nodes of $v_{r_{i}}$ within $t$ window size.

By making conditional independence assumption, the probability $\mathrm{P}(\{v_{r_{i-t}}, \cdots, v_{r_{i+t}}\}\setminus v_{r_{i}}|v_{r_{i}})$ can be expanded as
\begin{equation}
\mathrm{P}(\{v_{r_{i-t}}, \cdots, v_{r_{i+t}}\}\setminus v_{r_{i}}|v_{r_{i}}) = \prod_{j=i-t,j\neq i}^{i+t}\mathrm{P}(v_{r_{j}}|v_{r_{i}}).
\end{equation}

Following~\cite{zhang2017user}, considering all the generated random walks, the overall optimization problem of DeepWalk can be reformulated as
\begin{equation}
\min_{\mathrm{\Phi}} -\sum_{i=1}^{|\mathcal{V}|}\sum_{j=1}^{|\mathcal{V}|}n(v_{i},v_{j})\log\mathrm{P}(v_{j}|v_{i}),
\end{equation}where $n(v_{i},v_{j})$ is the total occurrence times of $v_{j}$ as $v_{i}$'s context nodes across all generated random walks within $t$ window size. The overall optimization problem can be solved by stochastically sampling a node context pair $(v_{i},v_{j})$ and minimizing the following partial objective:
\begin{equation} \label{deepwalk_partial_obj}
\mathcal{O}_{ij}^{s}=-\log\mathrm{P}(v_{j}|v_{i}).
\end{equation}

\section{SINE: Scalable Incomplete Network Embedding}
\label{sec-methodology}

\subsection{Model Architecture}

\begin{figure}
	\centering
	\begin{tikzpicture}
	[thick,scale=0.6, every node/.style={scale=0.6}]
	\matrix[nodes={red, draw, line width=0.2pt}, column sep=0.025cm](O1)
	{
		\node[circle] (O11) {\large\color{black}$\Sigma$} ; &
		\node[circle] (O12) {\large\color{black}$\Sigma$} ; &
		\node[draw=none,fill=none] {\large\color{black}$\cdots$}; &
		\node[circle] (O1j) {\large\color{black}$\Sigma$} ; &
		\node[draw=none,fill=none] {\large\color{black}$\cdots$}; &
		\node[circle] (O1V) {\large\color{black}$\Sigma$} ; \\
	};
	\matrix[right=0.4cm of O1, nodes={red, draw, line width=0.2pt}, column sep=0.025cm](O2)
	{
		\node[circle] (O21) {\large\color{black}$\Sigma$} ; &
		\node[circle] (O22) {\large\color{black}$\Sigma$} ; &
		\node[draw=none,fill=none] {\large\color{black}$\cdots$}; &
		\node[circle] (O2j) {\large\color{black}$\Sigma$} ; &
		\node[draw=none,fill=none] {\large\color{black}$\cdots$}; &
		\node[circle] (O2V) {\large\color{black}$\Sigma$} ; \\
	};
	\matrix[below left = 1cm and -1.45cm of O2, nodes={blue, draw, line width=0.2pt}, column sep=0.025cm](H)
	{
		\node[circle] (H1) {\large\color{black}$\Sigma$} ; &
		\node[circle] (H2) {\large\color{black}$\Sigma$} ; &
		\node[draw=none,fill=none] {\large\color{black}$\cdots$}; &
		\node[circle] (Hj) {\large\color{black}$\Sigma$} ; &
		\node[draw=none,fill=none] (Hdots) {\large\color{black}$\cdots$}; &
		\node[circle] (HV) {\large\color{black}$\Sigma$} ; \\
	};
	\matrix[below = 1cm of H, nodes={cyan, draw, line width=0.2pt}, column sep=0.025cm](I)
	{
		\node[rectangle] (I1) {\large\color{black}0} ; &
		\node[rectangle] (I2) {\large\color{black}0} ; &
		\node[draw=none,fill=none] {\large\color{black}$\cdots$}; &
		\node[rectangle,fill=cyan] (Ij) {\large\color{white}1} ; &
		\node[draw=none,fill=none] (Idots) {\large\color{black}$\cdots$}; &
		\node[rectangle] (IV) {\large\color{black}0} ; \\
	};
	
	\draw [purple!50,line width=1pt,-latex] (I) -- (H);
	\draw [purple!50,line width=1pt,-latex] (H) -- (O1);
	\draw [purple!50,line width=1pt,-latex] (H) -- (O2);
	
	\matrix[draw=none,fill=none, left = 0.1cm of O11, row sep=0.01cm](Out)
	{
		\node[text width=8em, text centered]{Output Layer};\\
		\node[text width=8em, text centered]{Softmax Classifiers};\\
	};
	
	\matrix[draw=none,fill=none, left = 0.1cm of H1, row sep=0.01cm](Hin)
	{
		\node[text width=8em, text centered]{Hidden Layer};\\
		\node[text width=8em, text centered]{Linear Neurons };\\
	};
	
	\matrix[draw=none,fill=none, left = 0.1cm of I1, row sep=0.01cm](In)
	{
		\node[text width=10em, text centered]{Input Layer};\\
		\node[text width=10em, text centered]{One-hot Representation};\\
	};
	
	\node[draw=none, fill=none, below = 0.02cm of Ij, text width=8em, text centered] {\small the position corresponding to $v_{i}$};
	
	\node[draw=none, fill=none, above = 0.02cm of O11, text width=8em, text centered] {\small $\mathrm{P}(v_{1}|v_{i})$};
	\node[draw=none, fill=none, above = 0.02cm of O1j, text width=8em, text centered]  {\small $\mathrm{P}(v_{j}|v_{i})$};
	\node[draw=none, fill=none, above = 0.02cm of O1V, text width=8em, text centered] {\small $\mathrm{P}(v_{|\mathcal{V}|}|v_{i})$};
	
	\node[draw=none, fill=none, above = 0.02cm of O21, text width=8em, text centered] {\small $\mathrm{P}(a_{1}|v_{i})$};
	\node[draw=none, fill=none, above = 0.02cm of O2j, text width=8em, text centered]  {\small $\mathrm{P}(a_{j}|v_{i})$};
	\node[draw=none, fill=none, above = 0.02cm of O2V, text width=8em, text centered] {\small $\mathrm{P}(a_{|\mathcal{A}|}|v_{i})$};
	
	\node[draw=none, fill=none, above left = 0.4cm and -0.9cm of Idots, text width=8em, text centered] {\small $W^{in}$};
	\node[draw=none, fill=none, above left = 0.4cm and -1.6cm of Hdots, text width=8em, text centered] {\small $W^{out,a}$};
	\node[draw=none, fill=none, above left = 0.4cm and 0.9cm of Hdots, text width=8em, text centered] {\small $W^{out,s}$};
	\end{tikzpicture}
	\caption{The model architecture of SINE. For each node $v_{i}$, SINE learns its representation by using it to predict its context node $v_{j}$ and its observable attribute $a_{j}$ so that nodes sharing similar context nodes or similar observed attributes are embedded closely in the new vector space. In this way, the incomplete structure and node attribute information is utilized flexibly to learn informative node representations.}
	\label{fig1:mechanism} 
\end{figure}
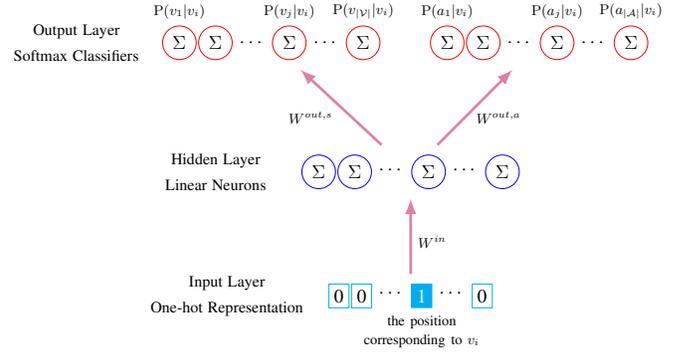

Inspired by DeepWalk~\cite{perozzi2014deepwalk}, SINE encodes network structure into node embeddings by allowing nodes sharing similar context nodes to have similar representations. This is achieved by minimizing the DeepWalk partial objective in Eq. (\ref{deepwalk_partial_obj}) for each node context pair $(v_{i},v_{j})$ collected from random walks.

In addition to topology structure, we also wish that nodes sharing similar attributes should be close in the new vector space. Following the mechanism of Skip-Gram~\cite{mikolov2013distributed}, we make the learned node representations respect this property by using node $v_{i}$ to predict its co-occurring attribute $a_{j}$ for each observed node attribute co-occurrence pair $(v_{i},a_{j})$. This is achieved by minimizing the following objective:

\begin{equation} \label{attribute_partial_obj}
\mathcal{O}_{ij}^{a}=-\log\mathrm{P}(a_{j}|v_{i}).
\end{equation}

As shown in Fig. \ref{fig1:mechanism}, the learning framework of SINE is a three-layer neural network: the first layer is the one-hot representation for each node $v_{i}$, the hidden layer is the node representation $\mathrm{\Phi}(v_{i})\in\mathbb{R}^{d}$ constructed by a linear transformation from the input layer with weight matrix $W^{in}\in\mathbb{R}^{|\mathcal{V}|\times d}$, the output layer is the softmax conditional probability $\mathrm{P}(v_{j}|v_{i})$ and $\mathrm{P}(a_{j}|v_{i})$, for each node $v_{j}$ and each attribute $a_{j}$, aggregated from the hidden layer with weight matrix $W^{out,s}\in\mathbb{R}^{d\times|\mathcal{V}|}$ and $W^{out,a}\in\mathbb{R}^{d\times|\mathcal{A}|}$, respectively.

Given the one-hot representation $\bm{p}^{(i)}\in\mathbb{R}^{|V|}$ of node $v_{i}$ with $\bm{p}^{(i)}_{i}=1$ and $\bm{p}^{(i)}_{j}=0$ for $j\neq i$, the node representation $\mathrm{\Phi}(v_{i})$ in the hidden layer is constructed as
\begin{equation}\label{hidden}
\mathrm{\Phi}(v_{i})={W^{in}}^{\mathrm{T}}\bm{p}^{(i)}=\bm{w}^{in}_{i},
\end{equation}where $\bm{w}^{in}_{i}$ is the transpose of the $i$-th row of $W^{in}\in\mathbb{R}^{|\mathcal{V}|\times d}$ (the weight matrix from the input layer to the hidden layer).

In the output layer, for node context pair $(v_{i},v_{j})$, the probability $\mathrm{P}(v_{j}|v_{i})$ is modeled by the softmax signal:
\begin{equation}\label{prob_context}
\mathrm{P}(v_{j}|v_{i})=\frac{\exp(\mathrm{\Phi}(v_{i})\cdot\bm{w}^{out,s}_{j})}{\sum_{k=1}^{|\mathcal{V}|}\exp(\mathrm{\Phi}(v_{i})\cdot\bm{w}^{out,s}_{k})},
\end{equation} where $\bm{w}^{out,s}_{j}$ is the $j$-th column of $W^{out,s}\in\mathbb{R}^{d\times|\mathcal{V}|}$ (the weight matrix from the hidden layer to the output layer for predicting node context).

Similarly, for the node attribute co-occurrence pair $(v_{i},a_{j})$, the probability $\mathrm{P}(a_{j}|v_{i})$ is modeled by
\begin{equation}\label{prob_content}
\mathrm{P}(a_{j}|v_{i})=\frac{\exp(\mathrm{\Phi}(v_{i})\cdot\bm{w}^{out,a}_{j})}{\sum_{k=1}^{|\mathcal{A}|}\exp(\mathrm{\Phi}(v_{i})\cdot\bm{w}^{out,a}_{k})},
\end{equation}where $\bm{w}^{out,a}_{j}$ is the $j$-th column of $W^{out,a}\in\mathbb{R}^{d\times|\mathcal{A}|}$ (the weight matrix from the hidden layer to the output layer for predicting node attribute).

By aggravating the structure preserving objective in Eq. (\ref{deepwalk_partial_obj}) and the node attribute preserving objective in Eq. (\ref{attribute_partial_obj}), node representations $\mathrm{\Phi}(\cdot)$ that well preserve the available structure and node content information, can be learned by solving the following overall optimization problem:
\begin{equation}\label{overall_optimization}
\min_{\mathrm{\Phi}}\mathcal{O},
\end{equation}where
\begin{equation}\label{overall_objective}
\begin{aligned}
\mathcal{O}=&-\alpha_{1}\sum_{i=1}^{|\mathcal{V}|}\sum_{j=1}^{|\mathcal{V}|}n(v_{i},v_{j})\mathrm{\log}\mathrm{P}(v_{j}|v_{i})\\
&-\alpha_{2}\sum_{i=1}^{|\mathcal{V}|}\sum_{j=1}^{|\mathcal{A}|}\Omega_{ij}X_{ij}\mathrm{\log}\mathrm{P}(a_{j}|v_{i}).
\end{aligned}
\end{equation}where $\alpha_{1}$ and $\alpha_{2}$ are the trade-off parameters that balance the structure preserving objective and the node content preserving objective. They are set as
\begin{equation*}
\alpha_{1}=\frac{1}{\sum_{i=1}^{|\mathcal{V}|}\sum_{j=1}^{|\mathcal{V}|}n(v_{i},v_{j})},\;\alpha_{2}=\frac{1}{\sum_{i=1}^{|\mathcal{V}|}\sum_{j=1}^{|\mathcal{A}|}\Omega_{ij}X_{ij}}.
\end{equation*}
In Eq. (\ref{overall_objective}), we only concern about the non-zero values of $n(v_{i},v_{j})$ and $\Omega_{ij}X_{ij}$, i.e., the node context pairs collected from observed links and the observed node attribute co-occurrence pairs, whose number is much smaller than $|\mathcal{V}|\times|\mathcal{V}|$ and $|\mathcal{V}|\times|\mathcal{A}|$, respectively. In this way, the available network structure and node content information can be fully utilized, and the negative impacts of missing information is diminished.

\subsection{Model Optimization}

We solve the overall optimization problem in Eq. (\ref{overall_optimization}) by minimizing the partial objective in Eq. (\ref{deepwalk_partial_obj}) and Eq. (\ref{attribute_partial_obj}) alternately with stochastic gradient descent by sampling a node context pair $(v_{i},v_{j})$ or a node attribute co-occurrence pair $(v_{i},a_{j})$ at each iteration, according to the distribution of $n(v_{i},v_{j})$, and $\Omega_{ij}X_{ij}$, respectively.

For the sampled node context pair $(v_{i},v_{j})$, negative sampling is used to speed up training. Thus, the partial objective $\mathcal{O}^{s}_{ij}$ in Eq. (\ref{deepwalk_partial_obj}) can be reformulates as
\begin{equation}
\begin{aligned}
\mathcal{O}^{s}_{ij} &= -\log\sigma(\mathrm{\Phi}(v_{i})\cdot\bm{w}_{j}^{out,s})\\
&-\sum_{k:v_{k}\in\mathcal{V}_{neg}}\log\sigma(-\mathrm{\Phi}(v_{i})\cdot\bm{w}_{k}^{out,s}),
\end{aligned}
\end{equation}where $\mathcal{V}_{neg}$ is the set of sampled negative nodes. The parameters are updated by gradient descent:
\begin{equation}\label{update_structure}
\left\lbrace
\begin{aligned}
&\bm{w}^{in}_{i} = \bm{w}^{in}_{i}  - \eta \frac{\partial\mathcal{O}^{s}_{ij}}{\partial\bm{w}^{in}_{i}},\\
&\bm{w}^{out,s}_{j} = \bm{w}^{out,s}_{j}  - \eta \frac{\partial\mathcal{O}^{s}_{ij}}{\partial\bm{w}^{out,s}_{j}},\\
&\bm{w}^{out,s}_{k} = \bm{w}^{out,s}_{k}  - \eta \frac{\partial\mathcal{O}^{s}_{ij}}{\partial\bm{w}^{out,s}_{k}},\; \mathrm{for} \; v_{k}\in\mathcal{V}_{neg},\\
\end{aligned}
\right.
\end{equation}where $\eta$ is the learning rate. The gradients are calculated as follows
\begin{equation}
\left\lbrace
\begin{aligned}
&\frac{\partial\mathcal{O}_{ij}^{s}}{\partial\bm{w}^{in}_{i}} =
 (\sigma(\mathrm{\Phi}(v_{i})\cdot\bm{w}_{j}^{out,s})-1)\bm{w}_{j}^{out,s}\\
&\quad\quad\;\;+\sum_{k:v_{k}\in\mathcal{V}_{neg}}\sigma(\mathrm{\Phi}(v_{i})\cdot\bm{w}_{k}^{out,s})\bm{w}_{k}^{out,s},\\
&\frac{\partial\mathcal{O}_{ij}^{s}}{\bm{w}^{out,s}_{j}} = (\sigma(\mathrm{\Phi}(v_{i})\cdot\bm{w}^{out,s}_{j})-1)\mathrm{\Phi}(v_{i}),\\
&\frac{\partial\mathcal{O}_{ij}^{s}}{\bm{w}^{out,s}_{k}} = \sigma(\mathrm{\Phi}(v_{i})\cdot\bm{w}^{out,s}_{k})\mathrm{\Phi}(v_{i}),\;\mathrm{for}\;v_{k}\in\mathcal{V}_{neg}.\\
\end{aligned}
\right.
\end{equation}

\begin{algorithm}[htb]
	\caption{SINE: Scalable Incomplete Network Embedding}
	\label{alg:SINE}
	\begin{algorithmic}[1]
		\REQUIRE ~~\\
		An incomplete network $G=(\mathcal{V},\mathcal{E},\mathcal{A},X,\Omega)$;
		\ENSURE ~~\\
		Node representation $\mathrm{\Phi}(\cdot)$ for each $v_{i}\in\mathcal{V}$;
		\STATE $\mathbb{S}$ $\leftarrow$ generate a set of random walks on $G$;
		\STATE $n(v_i,v_j)$ $\leftarrow$ count frequency of node context pairs ($v_{i},v_{j})$ in $\mathbb{S}$;
		\REPEAT
		\STATE draw a random number $r\in(0,1)$;
		\IF{$r\leqslant 0.5$}
		\STATE{$(v_i,v_j) \leftarrow$ sample a node context pair according to the distribution of $n(v_i,v_j)$;}
		\STATE{$\mathcal{V}_{neg}\leftarrow$ draw $K$ negative nodes;}
		\STATE{$(W^{in},W^{out,s}) \leftarrow$ update parameters with $(v_i,v_j,\mathcal{V}_{neg})$ and Eq.~(\ref{update_structure});}
		\ELSE
		\STATE{$(v_i,a_j) \leftarrow$ sample a node attribute pair according to the distribution of $\Omega_{ij}X_{ij}$;}
		\STATE{$\mathcal{A}_{neg}\leftarrow$ draw $K$ negative attributes;}
		\STATE{$(W^{in},W^{out,a}) \leftarrow$ update parameters with $(v_i,a_j,\mathcal{A}_{neg})$ and Eq.~(\ref{update_attribute});}
		\ENDIF
		\UNTIL {maximum number of iterations expire;}\label{code:iteration}
		\STATE construct node representation $\Phi(\cdot)$ with $W^{in}$ and Eq. (\ref{hidden});
		\STATE \textbf{return} $\mathrm{\Phi}(\cdot)$;
	\end{algorithmic}
\end{algorithm}

Similarly, for each sampled node attribute co-occurrence pair $(v_{i},a_{j})$, by adopting negative sampling, the partial objective $\mathcal{O}^{a}_{ij}$ in Eq. (\ref{attribute_partial_obj}) is approximated by
\begin{equation}
\begin{aligned}
\mathcal{O}^{a}_{ij} &= -\log\sigma(\mathrm{\Phi}(v_{i})\cdot\bm{w}_{j}^{out,a})\\
&-\sum_{k:v_{k}\in\mathcal{A}_{neg}}\log\sigma(-\mathrm{\Phi}(v_{i})\cdot\bm{w}_{k}^{out,a}),
\end{aligned}
\end{equation}where $\mathcal{A}_{neg}$ is the set of sampled negative attributes. The parameters are updated as
\begin{equation}\label{update_attribute}
\left\lbrace
\begin{aligned}
&\bm{w}^{in}_{i} = \bm{w}^{in}_{i}  - \eta \frac{\partial\mathcal{O}^{a}_{ij}}{\partial\bm{w}^{in}_{i}},\\
&\bm{w}^{out,a}_{j} = \bm{w}^{out,a}_{j}  - \eta \frac{\partial\mathcal{O}^{a}_{ij}}{\partial\bm{w}^{out,a}_{j}},\\
&\bm{w}^{out,a}_{k} = \bm{w}^{out,a}_{k}  - \eta \frac{\partial\mathcal{O}^{a}_{ij}}{\partial\bm{w}^{out,a}_{k}},\; \mathrm{for} \; a_{k}\in\mathcal{A}_{neg}.\\
\end{aligned}
\right.
\end{equation} The gradients are calculated as follows
\begin{equation}
\left\lbrace
\begin{aligned}
&\frac{\partial\mathcal{O}_{ij}^{a}}{\partial\bm{w}^{in}_{i}} =
(\sigma(\mathrm{\Phi}(v_{i})\cdot\bm{w}_{j}^{out,a})-1)\bm{w}_{j}^{out,a}\\
&\quad\quad\;\;+\sum_{k:a_{k}\in\mathcal{A}_{neg}}\sigma(\mathrm{\Phi}(v_{i})\cdot\bm{w}_{k}^{out,a})\bm{w}_{k}^{out,a},\\
&\frac{\partial\mathcal{O}_{ij}^{a}}{\bm{w}^{out,a}_{j}} = (\sigma(\mathrm{\Phi}(v_{i})\cdot\bm{w}^{out,a}_{j})-1)\mathrm{\Phi}(v_{i}),\\
&\frac{\partial\mathcal{O}_{ij}^{a}}{\bm{w}^{out,a}_{k}} = \sigma(\mathrm{\Phi}(v_{i})\cdot\bm{w}^{out,a}_{k})\mathrm{\Phi}(v_{i}),\;\mathrm{for}\;a_{k}\in\mathcal{A}_{neg}.\\
\end{aligned}
\right.
\end{equation}


\textbf{Algorithm}~\ref{alg:SINE} gives the major procedure of the proposed SINE algorithm. At step 1, SINE starts random walks with length $L$ at each node for $\gamma$ times, and at step 2, counts the frequency of node context pairs $n(v_{i},v_{j})$ within $t$ window size. At step 3-14, SINE updates the parameters with stochastic gradient descent. At each iteration, SINE samples a random switch variable $r\in(0,1)$ to determine whether the structure preserving objective or the node attribute preserving objective is to be minimized, with a threshold of 0.5. To sample node context pair and node attribute pair, the alias table method~\cite{li2014reducing} is used, which takes only $O(1)$ time. After the iteration is finished, node representations $\mathrm{\Phi}(\cdot)$ are constructed with $W^{in}$ and Eq. (\ref{hidden}).

The time complexity of SINE only relies on the maximum number of iterations and the dimension of learned node representations $d$. The maximum number of iterations is at the scale of $O(\max(nnz(\Omega_{ij}X_{ij}),|\mathcal{V}|))$, where $nnz(\Omega_{ij}X_{ij})$ is the number of non-zero values of $\Omega_{ij}X_{ij}$, i.e., the number of observed node attribute co-occurrence pairs, and $|\mathcal{V}|$ indicates the scale of node context pairs collected from random walks. Thus, SINE has an overall time complexity of $O(d\cdot\max(nnz(\Omega_{ij}X_{ij}),|\mathcal{V}|))$, ensuring its good scalability.

\section{Experiments}
\label{sec-experiments}

In this section, we present experimental results on real-world networks to verify the effectiveness and efficiency of the proposed SINE algorithm in learning informative node representations for incomplete networks.

\subsection{Benchmark Networks}

\renewcommand\arraystretch{1.15}
\begin{table}[t]
	\begin{center}
		\tabcolsep 6pt
		\caption{Summary of Four Real-world Networks}
		\scalebox{0.95}{
		\begin{tabular}{|c|c|c|c|c|}
			\hline
			& Cora & Citeseer & DBLP(Subgraph) & DBLP(Full)\\\hline
			$|\mathcal{V}|$ & 2,708 & 3,312 & 18,448 & 1,632,442 \\
			$|\mathcal{E}|$  & 5,278 & 4,732 & 45,611& 2,327,450 \\
			$|\mathcal{A}|$ & 1,433& 3,703 & 5,959 & 154,309 \\
			$nnz(X)$ & 49,216 & 105,165 & 108,016 &10,413,178 \\
			\# of Class & 7 & 6 & 4 & N/A\\\hline
		\end{tabular}}
		\label{dataset}
	\end{center}
\end{table}

Four real-world networks used in our experiments are detailed as follows:

\vspace{0.05cm}\noindent\textbf{Cora} and \textbf{Citeseer}\footnote{\url{https://linqs.soe.ucsc.edu/data}}: The Cora network contains 2,708 publications and 5,249 citations. The Citeseer network includes 3,312 publications and 4,732 citations. For Cora and Citeseer, each paper is represented by a 1,433-dimensional, and 3,703-dimensional binary vector, with each dimension indicating the presence/absence of the corresponding word.

\vspace{0.05cm}\noindent\textbf{DBLP(Subgraph)} and \textbf{DBLP(Full)}: The DBLP(Full) network is constructed by the papers and their citation relationships of the DBLP bibliographic network\footnote{\url{https://aminer.org/citation} (Version 3 is used)}. There are 1,632,442 papers and 2,327,450 citations in all. To construct the DBLP(Subgraph) network, we extract papers from the four research areas: \textit{Database}, \textit{Data Mining}, \textit{Artificial Intelligence}, \textit{Computer Vision}, according to papers' venue information and remove papers with no citations. The DBLP(Subgraph) network contains 18,448 papers and 45,661 citations. From paper titles, for DBLP(Subgraph) and DBLP(Full), we construct 5,959-dimensional and 154,309-dimensional binary node feature vectors, respectively, with each dimension indicating the presence/absence of the corresponding word.

For all networks, the link direction is ignored. The statistics of these networks are summarized in Table \ref{dataset}.

\subsection{Baseline Methods}

We compare SINE with the following baseline methods:

\vspace{0.05cm}\noindent\textbf{DeepWalk}~\cite{perozzi2014deepwalk} / \textbf{node2vec}~\cite{grover2016node2vec}: node2vec is equivalent to DeepWalk under the default setting $p=1$ and $q=1$. They learn node representations by preserving the similarity between nodes sharing similar contexts in random walks.

\vspace{0.05cm}\noindent\textbf{LINE-1}~\cite{tang2015line}: LINE-1 denotes the version of LINE that learns node representations by modeling the first-order proximity.

\vspace{0.05cm}\noindent\textbf{LINE-2}~\cite{tang2015line}: LINE-2 is the version of LINE that preserves the second-order proximity.

\vspace{0.05cm}\noindent\textbf{SDNE}~\cite{wang2016structural}: SDNE uses a semi-supervised autoencoder to learn deep node representations that preserve both the first-order and second-order proximity.

\vspace{0.05cm}\noindent\textbf{Attribute}: This baseline learns node representations from only node attributes with the SINE learning framework.

\vspace{0.05cm}\noindent\textbf{TADW}~\cite{yang2015network}: TADW incorporates node content features into DeepWalk's network representation learning paradigm via inductive matrix factorization~\cite{natarajan2014inductive}.

\vspace{0.05cm}\noindent\textbf{HSCA}~\cite{zhang2016homophily}: HSCA enhances TADW via enforcing a first-order proximity preserving objective.

\vspace{0.05cm}\noindent\textbf{UPP-SNE}~\cite{zhang2017user}: UPP-SNE learns node representations by performing a structure-aware non-linear mapping on node content features.

\vspace{0.05cm}\noindent\textbf{MVC-DNE}~\cite{yang2017properties}: MVC-DNE encodes network structure and node content attributes into node representations through the deep autoencoder based cross-view learning. The decoding based version is used.

Due to the large size of DBLP(Full), TADW, HSCA, UPP-SNE and MVC-DNE can not run on this dataset. Thus, on DBLP(full), only DeepWalk, LINE-1, LINE-2 and Attribute are compared with SINE in Section~\ref{subsection: runningtime}.

\subsection{Experimental Settings}

For DeepWalk, UPP-SNE and SINE, we set the length of random walks $L=100$, the number of random walks starting from each node $\gamma=40$, and the window size $t=10$.

For fair comparison, DeepWalk and UPP-SNE are trained using the same strategy with SINE: we first collect node context pairs, and then update parameters with stochastic gradient descent by sampling a node context pair at each iteration. Negative sampling is adopted by DeepWalk, LINE-1, LINE-2, UPP-SNE, Attribute, and SINE, where the number of negative samples $K$ is set to 5 uniformly. For the 6 stochastic gradient descent based algorithms, the maximum number of iterations $I$ is set to 100 million on Cora, Citeseer and DBLP(Subgraph), and 1 billion on DBLP(Full), and the learning rate $\eta$ decreases from the starting value $\eta_{0}=0.025$ to $\eta_{\tau}=\eta_{0}(1-\tau/I)$ after every 10,000 iterations, with $\tau$ being the number of elapsed iterations.  Parameters of TADW and HSCA are set to their default values. For SDNE, its hyperparameters $\alpha$ and $\nu$ are set to 0.01, and $\beta$ is set to 10. The number of neurons at each layer is set to 2708-256, 3312-256, and 18,448-1,024-256 on Cora, Citeseer, and DBLP(Subgraph), respectively. For MVC-DNE, on Cora, Citeseer, and DBLP(Subgraph), the number of neurons at each layer in structure view is respectively set to 2708-128, 3312-128, and 18,448-512-128, and the number of neurons at each layer in node attribute view is set to 1,433-128, 3,703-128, and 5,959-128. For SDNE and MVC-DNE, 500 epochs are run for both pre-training and parameter fine-tuning. Other parameters of SDNE and MVC-DNE are set according to~\cite{yang2017properties}. For SINE and all baseline methods, the dimension of learned node representations is set to 256.


\subsection{Performance Comparison on Incomplete Networks}
In this section, we conduct experiments to compare the performance of SINE and baseline methods on incomplete networks. To better understand the ability of different network embedding algorithms to deal with missing data, we investigate the following four research questions:

\begin{center}
\begin{tabular}{|c|L{7.2cm}|}
\hline
Q1 & How is the performance affected when a portion of node attributes are missing compared with the complete network? (\textbf{complete attribute vs. incomplete attribute}) \\\hline

Q2 & How does the performance change when the attributes of structurally important nodes are missing compared with missing attributes for randomly selected nodes? (\textbf{random vs. important $X$ row missing}) \\\hline

Q3 & How does the performance change when the attributes at important dimensions are missing compared with missing attributes at random dimensions? (\textbf{random vs. important $X$ column missing}) \\\hline

Q4 & How do different network embedding algorithms perform for link prediction when a portion of edges are missing? \\\hline
\end{tabular}
\end{center}

To answer research questions Q1, Q2, and Q3, we compare the performance of different network embedding algorithms on Cora, Citeseer and DBLP(Subgraph) under 5 settings: (1) complete network, (2) randomly selecting 50\% nodes and dropping all of their attributes (random $X$ row missing), (3) selecting the top 50\% structurally important nodes measured by degree and dropping their attributes (important $X$ row missing), (4) missing node attributes at randomly selected 50\% dimensions (random $X$ column missing), and (5) missing node attributes at top 50\% important dimensions measured by mutual information with class label (important $X$ column missing). As the baselines of Attribute, TADW, HSCA and UPP-SNE require all nodes to have observed attributes, under settings (2) and (3), for nodes with no attributes, we fill their attribute values with the observed modes at each dimension. To compare the performance, using the learned node representations as features, we conduct multi-class node classification experiments on Cora, Citeseer and DBLP, and carry out node clustering experiments on Cora.

To answer research question Q4, we perform link prediction experiments on Cora and DBLP(Full). We randomly remove 30\%, 50\% and 70\% links, learn node representations from the respective incomplete networks with different algorithms, and compare their performance for predicting missing links.

\renewcommand\arraystretch{1}
\begin{table*}[t] 
	\begin{minipage}[b]{0.5\textwidth} 
			\centering
			\scriptsize
			\tabcolsep 5pt
			\caption{Node Classification Results on Cora}
			\scalebox{0.85}{
				\begin{tabular}{|C{1cm}|L{1.25cm}|C{1cm}|C{1cm}|C{1cm}|C{1cm}|C{1cm}|}
					\hline
					& \multirow{3}{*}{Method} & \multirow{3}{*}{Complete} & \multicolumn{4}{c|}{Incomplete} \\\cline{4-7}
					& & & \multicolumn{2}{c|}{$X$ Row Missing} & \multicolumn{2}{c|}{$X$ Column Missing} \\\cline{4-7}
					& & & Random & Important & Random & Important \\\hline
					\multirow{10}{*}{Micro-$F_{1}$} & DeepWalk & 0.8245 & \underline{0.8245} & \underline{0.8245} & 0.8245 & \textbf{0.8245} \\
					& LINE-1 & 0.7697 & 0.7697 & 0.7697 & 0.7697 & 0.7697 \\
					& LINE-2 & 0.7081 & 0.7081 & 0.7081 & 0.7081 & 0.7081 \\
					& SDNE & 0.6047 & 0.6047 & 0.6047 & 0.6047 & 0.6047 \\
					& Attribute & 0.7222 & 0.4905 & 0.4721 & 0.6581 & 0.2592 \\
					& TADW & \textbf{0.8543} & 0.7387 & 0.3982 & \underline{0.8383} & 0.7068 \\
					& HSCA & \textbf{0.8569} & 0.7366 & 0.3791 & \textbf{0.8530} & 0.7309 \\
					& UPP-SNE & 0.8191 & 0.5538 & 0.5438 & 0.8058 & 0.7201  \\
					& MVC-DNE & 0.7528 & 0.6258 & 0.6500 & 0.7075 & 0.4978 \\
					& SINE & \underline{0.8340} & \textbf{0.8329} & \textbf{0.8383} & 0.8332 & \underline{0.8010} \\\hline
					\multirow{10}{*}{Macro-$F_{1}$} & DeepWalk & 0.8184 & \underline{0.8184} & \underline{0.8184} & 0.8184 & \textbf{0.8184} \\
					& LINE-1 & 0.7673 & 0.7673 & 0.7673 & 0.7673 & 0.7673 \\
					& LINE-2 & 0.6986 & 0.6986 & 0.6986 & 0.6986 & 0.6986 \\
					& SDNE & 0.5811 & 0.5811 & 0.5811 &  0.5811 & 0.5811 \\
					& Attribute & 0.6927 & 0.4373 & 0.4130 & 0.6302 & 0.1427 \\
					& TADW & \textbf{0.8457} & 0.7297 & 0.2903 & \underline{0.8295} & 0.6890 \\
					& HSCA & \textbf{0.8487} & 0.7151 & 0.2521 & \textbf{0.8445} & 0.7130 \\
					& UPP-SNE & 0.8088 & 0.5322 & 0.5227 & 0.7985 & 0.7072 \\
					& MVC-DNE & 0.7263 & 0.5872 & 0.6176 & 0.6808 & 0.4447 \\
					& SINE & \underline{0.8214} & \textbf{0.8218} & \textbf{0.8266} & 0.8228 & \underline{0.7896} \\\hline
			\end{tabular}}
			\label{Classification_Res_Cora}
	\end{minipage}%
	\begin{minipage}[b]{0.5\textwidth} 
		   \centering
			\scriptsize
			\tabcolsep 5pt
			\caption{Node Classification Results on Citeseer}
			\scalebox{0.85}{
				\begin{tabular}{|C{1cm}|L{1.25cm}|C{1cm}|C{1cm}|C{1cm}|C{1cm}|C{1cm}|}
					\hline
					& \multirow{3}{*}{Method} & \multirow{3}{*}{Complete} & \multicolumn{4}{c|}{Incomplete} \\\cline{4-7}
					& & & \multicolumn{2}{c|}{$X$ Row Missing} & \multicolumn{2}{c|}{$X$ Column Missing} \\\cline{4-7}
					& & & Random & Important & Random & Important \\\hline
					\multirow{10}{*}{Micro-$F_{1}$} & DeepWalk & 0.6038 & \underline{0.6038} & \underline{0.6038} & 0.6038 & \textbf{0.6038} \\
					& LINE-1 & 0.5684 & 0.5684 & 0.5684 & 0.5684 & \underline{0.5684} \\
					& LINE-2 & 0.4673 & 0.4673 & 0.4673 & 0.4673 & 0.4673 \\
					& SDNE & 0.4614 & 0.4614 & 0.4614 & 0.4614 & 0.4614 \\
					& Attribute & 0.6883 & 0.4325 & 0.4287 & 0.6549 & 0.2264 \\
					& TADW & \underline{0.6957} & 0.5086 & 0.3267 & \textbf{0.7014} & 0.5199 \\
					& HSCA & 0.6825 & 0.3708 & 0.3128 & \textbf{0.7031} & 0.5358 \\
					& UPP-SNE & \textbf{0.7124} & 0.4418 & 0.4551 & \underline{0.6858} & \underline{0.5662} \\
					& MVC-DNE & \underline{0.6954} & 0.5642 & 0.5467 & 0.6578 & 0.3187 \\
					& SINE & \textbf{0.7136} & \textbf{0.6791} & \textbf{0.6882} & \textbf{0.7030} & \underline{0.5682} \\\hline
					\multirow{10}{*}{Macro-$F_{1}$} & DeepWalk & 0.5465 & \underline{0.5465} & \underline{0.5465} & 0.5465 & \textbf{0.5465} \\
					& LINE-1 & 0.5300 & 0.5300 & 0.5300 & 0.5300 & \underline{0.5300} \\
					& LINE-2 & 0.4192 & 0.4192 & 0.4192 & 0.4192 & 0.4192 \\
					& SDNE & 0.3986 & 0.3986 & 0.3986  & 0.3986 & 0.3986 \\
					& Attribute & 0.6317 & 0.4051 & 0.4018 & 0.5951 & 0.1962 \\
					& TADW & \underline{0.6403} & 0.4673 & 0.2806 & \textbf{0.6480} & 0.4613 \\
					& HSCA & 0.6292 & 0.3191 & 0.2589 & \textbf{0.6505} & 0.4740 \\
					& UPP-SNE & \textbf{0.6567} & 0.4135 & 0.4323 & 0.6230 & 0.5027 \\
					& MVC-DNE & 0.6269 & 0.4996 & 0.4788  & 0.5851 & 0.2714 \\
					& SINE & \textbf{0.6564} & \textbf{0.6201} & \textbf{0.6256} & \underline{0.6401} & 0.5017 \\\hline
			\end{tabular}}
			\label{Classification_Res_Citeseer}
	\end{minipage} 
\end{table*}

\begin{table*}[t] 
	\begin{minipage}[b]{0.5\textwidth} 
		\centering
		\scriptsize
		\tabcolsep 5pt
		\caption{Node Classification Results on DBLP(Subgraph)}
		\scalebox{0.85}{
			\begin{tabular}{|C{1cm}|L{1.25cm}|C{1cm}|C{1cm}|C{1cm}|C{1cm}|C{1cm}|}
				\hline
				& \multirow{3}{*}{Method} & \multirow{3}{*}{Complete} & \multicolumn{4}{c|}{Incomplete} \\\cline{4-7}
				& & & \multicolumn{2}{c|}{$X$ Row Missing} & \multicolumn{2}{c|}{$X$ Column Missing} \\\cline{4-7}
				& & & Random & Important & Random & Important \\\hline
				\multirow{10}{*}{Micro-$F_{1}$} & DeepWalk & 0.8005 & \underline{0.8005} & \underline{0.8005} & \underline{0.8005} & \textbf{0.8005} \\
				& LINE-1 & 0.7810 & 0.7810 & 0.7810 & 0.7810 & 0.7810 \\
				& LINE-2 & 0.7125 & 0.7125 & 0.7125 & 0.7125 & 0.7125 \\
				& SDNE & 0.6464 & 0.6464 & 0.6464 & 0.6464 & 0.6464 \\
				& Attribute & 0.7509 & 0.5665 & 0.6275 & 0.6226 & 0.3952 \\
				& TADW & 0.8070 & 0.7195 & 0.4544 & 0.7533 & 0.4928 \\
				& HSCA & 0.8023 & 0.6227 & 0.4178 & 0.7532 & 0.4628 \\
				& UPP-SNE & \underline{0.8267} & 0.6092 & 0.6606 & 0.7078 & 0.4141 \\
				& MVC-DNE & 0.7697 & 0.6693 & 0.7089 & 0.7025 & 0.6232 \\
				& SINE & \textbf{0.8370} & \textbf{0.8278} & \textbf{0.8370} & \textbf{0.8275} & \underline{0.7953} \\\hline
				\multirow{10}{*}{Macro-$F_{1}$} & DeepWalk & 0.7176 & \underline{0.7176} & \underline{0.7176} & \underline{0.7176} & \textbf{0.7176} \\
				& LINE-1 & 0.6896 & 0.6896 & 0.6896 & 0.6896 & 0.6896 \\
				& LINE-2 & 0.6129 & 0.6129 & 0.6129 & 0.6129 & 0.6129 \\
				& SDNE & 0.3973 & 0.3973 & 0.3973 & 0.3973 & 0.3973 \\
				& Attribute & 0.6604 & 0.4403 & 0.5412 & 0.4870 & 0.1882 \\
				& TADW & 0.7271 & 0.5539 & 0.2935 & 0.6443 & 0.2760 \\
				& HSCA & 0.6926 & 0.4537 & 0.2436 & 0.6140 & 0.2549 \\
				& UPP-SNE & \underline{0.7668} & 0.5196 & 0.5988 & 0.6187 & 0.1985 \\
				& MVC-DNE & 0.6699 & 0.5222 & 0.5961 & 0.5854 & 0.4022 \\
				& SINE & \textbf{0.7731} & \textbf{0.7620} & \textbf{0.7727} & \textbf{0.7618} & \underline{0.7122} \\\hline
		\end{tabular}}
		\label{Classification_Res_DBLP}
	\end{minipage}%
	\begin{minipage}[b]{0.5\textwidth} 
			\centering
		\scriptsize
		\tabcolsep 5pt
		\caption{Node Clustering Results on Cora}
		\scalebox{0.85}{
			\begin{tabular}{|C{1cm}|L{1.25cm}|C{1cm}|C{1cm}|C{1cm}|C{1cm}|C{1cm}|}
				\hline
				& \multirow{3}{*}{Method} & \multirow{3}{*}{Complete} & \multicolumn{4}{c|}{Incomplete} \\\cline{4-7}
				& & & \multicolumn{2}{c|}{$X$ Row Missing} & \multicolumn{2}{c|}{$X$ Column Missing} \\\cline{4-7}
				& & & Random & Important & Random & Important \\\hline
				\multirow{10}{*}{Accuracy} & DeepWalk & 0.6097 & \underline{0.6097} & \underline{0.6097} & \underline{0.6097} & \underline{0.6097} \\
				& LINE-1 & 0.3444 & 0.3444 & 0.3444 & 0.3444 & 0.3444 \\
				& LINE-2 & 0.4162 & 0.4162 & 0.4162 & 0.4162 & 0.4162 \\
				& SDNE & 0.3897 & 0.3897 & 0.3897 & 0.3897 & 0.3897 \\
				& Attribute & 0.3869 & 0.3216 & 0.3240 & 0.3454 & 0.3021 \\
				& TADW & 0.3561 & 0.3123 & 0.3045 & 0.4000 & 0.3316 \\
				& HSCA & 0.3721 & 0.3257 & 0.3046 & 0.3956 & 0.3357 \\
				& UPP-SNE & \underline{0.6270} & 0.4496 & 0.4269 & \textbf{0.6319} & 0.5573 \\
				& MVC-DNE & 0.6228 & 0.4058 & 0.3685 & 0.5352 & 0.3021 \\
				& SINE & \textbf{0.6323} & \textbf{0.6397} & \textbf{0.6355} & \textbf{0.6343} & \textbf{0.6297} \\\hline
				\multirow{10}{*}{NMI} & DeepWalk & 0.4165 & \underline{0.4165} & \underline{0.4165} & 0.4165 & \underline{0.4165} \\
				& LINE-1 & 0.0910 & 0.0910 & 0.0910 & 0.0910 & 0.0910 \\
				& LINE-2 & 0.1806 & 0.1806 & 0.1806 & 0.1806 & 0.1806 \\
				& SDNE & 0.1409 & 0.1409  & 0.1409 & 0.1409 & 0.1409 \\
				& Attribute & 0.1475 & 0.0528 & 0.0661 & 0.0836 & 0.0056 \\
				& TADW & 0.1377 & 0.0321 & 0.0120 & 0.1603 & 0.0698 \\
				& HSCA & 0.1646 & 0.0556 & 0.0113 & 0.1892 & 0.0767 \\
				& UPP-SNE & \underline{0.4377} & 0.2171 & 0.2003 & \underline{0.4427} & 0.3490 \\
				& MVC-DNE & 0.3737 & 0.1326 & 0.1119 & 0.3018 & 0.0067 \\
				& SINE & \textbf{0.4456} & \textbf{0.4458} & \textbf{0.4401} & \textbf{0.4468} & \textbf{0.4376} \\\hline
		\end{tabular}}
		\label{Clustering_Res_Cora}
	\end{minipage} 
\end{table*}

\subsubsection{Comparison of Classification Performance}
Using the learned node representations as features, we train an SVM classifier (with the LIBLINEAR implementation~\cite{fan2008liblinear}) on the randomly selected 50\% samples,  and then classify the remainder 50\% samples with the learned classifier. The random training and test data split is repeated for 10 times, and averaged Micro-$F_{1}$ and Macro-$F_{1}$ values are used to evaluate the classification performance.

Tables~\ref{Classification_Res_Cora}-\ref{Classification_Res_DBLP} report node classification results of different network embedding algorithms on Cora, Citeseer and DBLP(Full), under the five settings: (1) complete, (2) random $X$ row missing, (3) important $X$ row missing, (4) random $X$ column missing, and (5) important $X$ column missing. For each setting, the best Micro-$F_{1}$ and Macro-$F_{1}$ values are \textbf{bold-faced}, and the second best performers are \underline{underlined}.

By comparing column 3 with columns 4-7 in Tables \ref{Classification_Res_Cora}-\ref{Classification_Res_DBLP}, we can respond to research question Q1: when node attributes become incomplete, the performance of the existing attributed network embedding baselines (Attribute, TADW, HSCA, UPP-SNE and MVC-DNE) drops remarkably in most cases, while SINE often shows greater stability. This attributes to the flexible way that SINE leverages node attributes, making it able to best utilize observed node attributes and to diminish the negative impact caused by missing attributes.

To better understand research question Q2, we compare column 4 with column 5 in Tables \ref{Classification_Res_Cora}-\ref{Classification_Res_DBLP}. When the attributes of structurally important nodes are missing, the performance of TADW and HSCA degrades dramatically on all three datasets. Interestingly, on DBLP, the Attribute baseline experiences a performance gain. This might be due to the fact that structurally important nodes tend to have strong correlations with neighbor nodes on attribute values, resulting in information redundancy. When such redundancy is removed, Attribute achieves better classification performance. Due to the same reason, UPP-SNE and MVC-DNE share the same trends with Attribute on DBLP. However, under both settings, SINE exhibits more stable performance and outperforms all other baselines.

\vspace*{-0.05cm}
To answer research question Q3, we compare column 6 with column 7 in Tables \ref{Classification_Res_Cora}-\ref{Classification_Res_DBLP}. The performance of Attribute decreases dramatically when node attributes at important dimensions are missing, compared with the missing in randomly selected dimensions. Accordingly, attributed network embedding algorithms consistently experience a dramatic performance drop, to a level inferior to the only structure preserving network embedding algorithms. In this case, the remaining node attributes are of poor quality, making them deteriorate rather than complement network structure in learning network embeddings. By contrast, the performance of SINE drops less significantly, demonstrating its better robustness to missing node attributes. 

From Tables \ref{Classification_Res_Cora}-\ref{Classification_Res_DBLP}, we can conclude that SINE achieves the best overall performance under all node attribute missing settings. This not only verifies the effectiveness of SINE in handling missing node attributes, but also signifies its great potential to solve real-world applications with missing data.

\subsubsection{Clustering Performance Comparison}

As a complement to node classification, we also conduct node clustering experiments on Cora. We feed node representations learned by different network embedding algorithms into the $K$-means clustering algorithm and group them into 7 categories. To alleviate the impact caused by random initialization, we run $K$-means for 20 times and report averaged Accuracy and NMI~\cite{strehl2002cluster} values. The clustering results are presented in Table \ref{Clustering_Res_Cora}, with the best and second performer highlighted by \textbf{bold} and \underline{underline} respectively. Similar to node classification results, SINE achieves the best clustering performance with small variance across all node attribute missing settings.

\subsubsection{Link Prediction Performance Comparison}

To answer research question Q4, we carry out link prediction experiments on Cora and DBLP(Full). Specifically, we compare the performance of different network embedding algorithms on link prediction, when a portion of edges are missing. Following~\cite{grover2016node2vec}, we perform link prediction using the edge features constructed from the learned node representations with the operators listed in Table~\ref{edge_feature}. We randomly remove 30\%, 50\% and 70\% of edges. To construct the test set, for each removed edge $(v_{i},v_{j})$, we randomly sample a negative node pair $(v_{i},v_{k})$ with $(v_{i},v_{k})\notin\mathcal{E}$ as negative ground truth.  To construct the training set, for each connected node pair $(v_{i},v_{j})$, we randomly sample a negative node pair $(v_{i},v_{k})$, with no edges between $v_{i}$ and $v_{k}$ observed in the remaining network.

SVM implemented by LIBLINEAR~\cite{fan2008liblinear} is used to perform training and testing on edge features. Due to the memory limitation, for DBLP(Full), 5\% training  and test samples are randomly selected from the two original sets respectively. We also compare SINE with the common neighbor based heuristic link prediction methods, which are given in Table~\ref{heuristic_link_prediction}. The Area Under Curve (AUC) score is used to evaluate the link prediction performance.

\renewcommand\arraystretch{1.0}
\begin{table}[t]
	\centering
	\caption{Operators to construct edge features}
	\scalebox{0.95}{
		\begin{tabular}{|c|c|c|}
			\hline
			Operator & Symbol & Definition\\\hline
			Average & $\boxplus$ & $\left[\mathrm{\Phi}(v_{i})\boxplus\mathrm{\Phi}(v_{j})\right]_{k}=\frac{\mathrm{\Phi}_{k}(v_{i})+\mathrm{\Phi}_{k}(v_{j})}{2}$ \\
			Hadamard & $\boxdot$ & $\left[\mathrm{\Phi}(v_{i})\boxdot\mathrm{\Phi}(v_{j})\right]_{k}=\mathrm{\Phi}_{k}(v_{i})\cdot\mathrm{\Phi}_{k}(v_{j})$\\
			Weighted-L1 & ${\lVert\cdot\rVert}_{\bar{1}}$ & ${\lVert\mathrm{\Phi}(v_{i})\cdot\mathrm{\Phi}(v_{j})\rVert}_{\bar{1}k}=|\mathrm{\Phi}_{k}(v_{i})-\mathrm{\Phi}_{k}(v_{j})|$\\
			Weighted-L2 & ${\lVert\cdot\rVert}_{\bar{2}}$ & ${\lVert\mathrm{\Phi}(v_{i})\cdot\mathrm{\Phi}(v_{j})\rVert}_{\bar{2}k}=(\mathrm{\Phi}_{k}(v_{i})-\mathrm{\Phi}_{k}(v_{j}))^{2}$\\\hline
	\end{tabular}}
	\label{edge_feature}
\end{table}

\begin{table}[t]
	\centering
	\caption{Heuristic scores for predicting the link between node pair $(v_{i},v_{j})$ with their direct neighbor sets $\mathcal{N}(v_{i})$ and $\mathcal{N}(v_{j})$}
	\scalebox{0.95}{
		\begin{tabular}{|l|c|}
			\hline
			Score & Definition\\\hline
			Common Neighbors & $|\mathcal{N}(v_{i})\cap\mathcal{N}(v_{j})|$\\
			Jaccard's Coefficient & $\frac{|\mathcal{N}(v_{i})\cap\mathcal{N}(v_{j})|}{|\mathcal{N}(v_{i})\cup\mathcal{N}(v_{j})|}$\\
			Adamic-Adar Score & $\sum_{v_{k}\in\mathcal{N}(v_{i})\cap\mathcal{N}(v_{j})}\frac{1}{\log|\mathcal{N}(v_{k})|}$\\
			Preferential Attachment & $|\mathcal{N}(v_{i})|\cdot|\mathcal{N}(v_{j})|$\\\hline
	\end{tabular}}
	\label{heuristic_link_prediction}
\end{table}

\begin{table}[t]
	\centering
	\scriptsize
	\tabcolsep 10pt
	\caption{AUC Values for Link Prediction on Cora}
	\scalebox{0.9}{
	\begin{tabular}{|c|l|c|c|c|}
		\hline
		Operator & Method & 30\% & 50\%  & 70\% \\ \hline
		& Common Neighbors & 0.5115 & 0.5045 & 0.5019 \\
		& Jaccard's Coefficient & 0.5115 & 0.5045 & 0.5019 \\
		& Adamic-Adar & 0.5115 & 0.5045 & 0.5019 \\
		& Pref. Attachment & 0.5445 & 0.5282 & 0.5229 \\ \hline
		\multirow{10}{*}{Average}& DeepWalk & 0.5123 & 0.4952 & 0.5039 \\
		& LINE-1 & 0.5542 & 0.5307 & 0.5157 \\
		& LINE-2 & 0.5280 & 0.5220 & 0.5178 \\
		& SDNE & 0.5585 & 0.5488 & 0.5384 \\
		& Attribute & 0.5320 & 0.5326 & 0.5264 \\
		& TADW & 0.5571 & 0.5434 & 0.5386 \\
		& HSCA & 0.5543 & 0.5368 & 0.5272 \\
		& UPP-SNE & 0.5336 & 0.5213 & 0.5214 \\
		& MVC-DNE & 0.5348  & 0.5202 & 0.5269 \\
		& SINE &0.5507 & 0.5425 & 0.5375 \\\hline
		\multirow{10}{*}{Hadamard} & DeepWalk & 0.7632 & 0.6657 & 0.5563 \\
		& LINE-1 & 0.6702 & 0.6003 & 0.5403 \\
		& LINE-2 & 0.6763 & 0.5808 & 0.5255 \\
		& SDNE & 0.5730 & 0.5516 & 0.5293 \\
		& Attribute & 0.8095 & 0.7966 & \underline{0.7989} \\
		& TADW & 0.8361 & 0.7850 & 0.7202 \\
		& HSCA & \underline{0.8623} & 0.8108 & 0.7351 \\
		& UPP-SNE & \underline{0.8615} & \underline{0.8160} & 0.7415 \\
		& MVC-DNE & 0.6749 & 0.6090 & 0.5643 \\
		& SINE & \textbf{0.8804} & \textbf{0.8545} & \textbf{0.8289} \\\hline
		\multirow{10}{*}{Weighted-L1} & DeepWalk & 0.8368 & 0.7302 & 0.6015 \\
		& LINE-1 & 0.6303 & 0.5900 & 0.5199 \\
		& LINE-2 & 0.6429 & 0.5726 & 0.5109 \\
		& SDNE & 0.5643 & 0.5365 & 0.5219 \\
		& Attribute & 0.7402 & 0.7212 & 0.7179 \\
		& TADW & 0.7443 & 0.6611 & 0.5834 \\
		& HSCA & 0.7725 & 0.6840 & 0.5859 \\
		& UPP-SNE & 0.8512 & 0.7970 & 0.7095 \\
		& MVC-DNE & 0.7883 & 0.7662 & 0.7418 \\
		& SINE & 0.8438 & 0.8035 & 0.7643 \\\hline
		\multirow{10}{*}{Weighted-L2} & DeepWalk & 0.8368 & 0.7356 & 0.6041 \\
		& LINE-1 & 0.6102 & 0.5720 & 0.5131 \\
		& LINE-2 & 0.6631 & 0.4888 & 0.4874 \\
		& SDNE & 0.5267 & 0.4915 & 0.5198 \\
		& Attribute & 0.7398 & 0.7311 & 0.7333 \\
		& TADW & 0.7279 & 0.6540 & 0.5742 \\
		& HSCA & 0.7400 & 0.6630 & 0.5749 \\
		& UPP-SNE & 0.8414 & 0.7937 & 0.7149 \\
		& MVC-DNE & 0.7359 & 0.7331 & 0.7264 \\
		& SINE & 0.8344 & 0.7965 & 0.7581 \\\hline
	\end{tabular}}
	\label{LinkPrediction_Res_Cora}
\end{table}

\begin{table}[t]
	\centering
	\scriptsize
	\tabcolsep 10pt
	\caption{AUC Values for Link Prediction on DBLP(Full)}
	\scalebox{0.9}{
		\begin{tabular}{|c|l|c|c|c|}
			\hline
			Operator & Method & 30\% & 50\%  & 70\% \\ \hline
			& Common Neighbors & 0.5024 & 0.5013 & 0.5006 \\
			& Jaccard's Coefficient & 0.5024 & 0.5013 & 0.5006 \\
			& Adamic-Adar & 0.5024 & 0.5013 & 0.5006 \\
			& Pref. Attachment & 0.8898 & 0.8768 & 0.8513 \\ \hline
			\multirow{5}{*}{Average}& DeepWalk & 0.9016 & 0.8896 & 0.8619 \\
			& LINE-1 & 0.5500 & 0.5389 & 0.5281 \\
			& LINE-2 & 0.8810 & 0.8689 & 0.8424 \\
			& Attribute & 0.6215 & 0.6205 & 0.6185 \\
			& SINE & 0.8747 & 0.8692 & 0.8533 \\\hline
			\multirow{5}{*}{Hadamard} & DeepWalk & \underline{0.9841} & \underline{0.9668} & \underline{0.9270} \\
			& LINE-1 & \underline{0.9835} & 0.9590 & 0.8999 \\
			& LINE-2 & 0.9766 & 0.9576 & 0.9146 \\
			& Attribute & 0.8490 & 0.8500 & 0.8507 \\
			& SINE & \textbf{0.9887} & \textbf{0.9831} & \textbf{0.9670} \\\hline
			\multirow{5}{*}{Weighted-L1} & DeepWalk & 0.9766 & 0.9463 & 0.8819 \\
			& LINE-1 & 0.8157 & 0.7573 & 0.6397 \\
			& LINE-2 & 0.8787 & 0.8601 & 0.8400 \\
			& Attribute & 0.7818 & 0.7819 & 0.7778 \\
			& SINE & 0.9382& 0.9150 & 0.8497 \\\hline
			\multirow{5}{*}{Weighted-L2} & DeepWalk & 0.9390 & 0.9018 & 0.8160 \\
			& LINE-1 & 0.8055 & 0.7452 & 0.6172 \\
			& LINE-2 & 0.6910 & 0.6618 & 0.8246 \\
			& Attribute & 0.7681 & 0.7701 & 0.7688 \\
			& SINE & 0.9029 & 0.8787 & 0.8113 \\\hline
	\end{tabular}}
	\label{LinkPrediction_Res_DBLPFull}
\end{table}

\begin{figure*}[t]
	\centering
	\subfigure[$I$]{
		\label{fig:para:subfig:iteration}
		\includegraphics[width=2in]{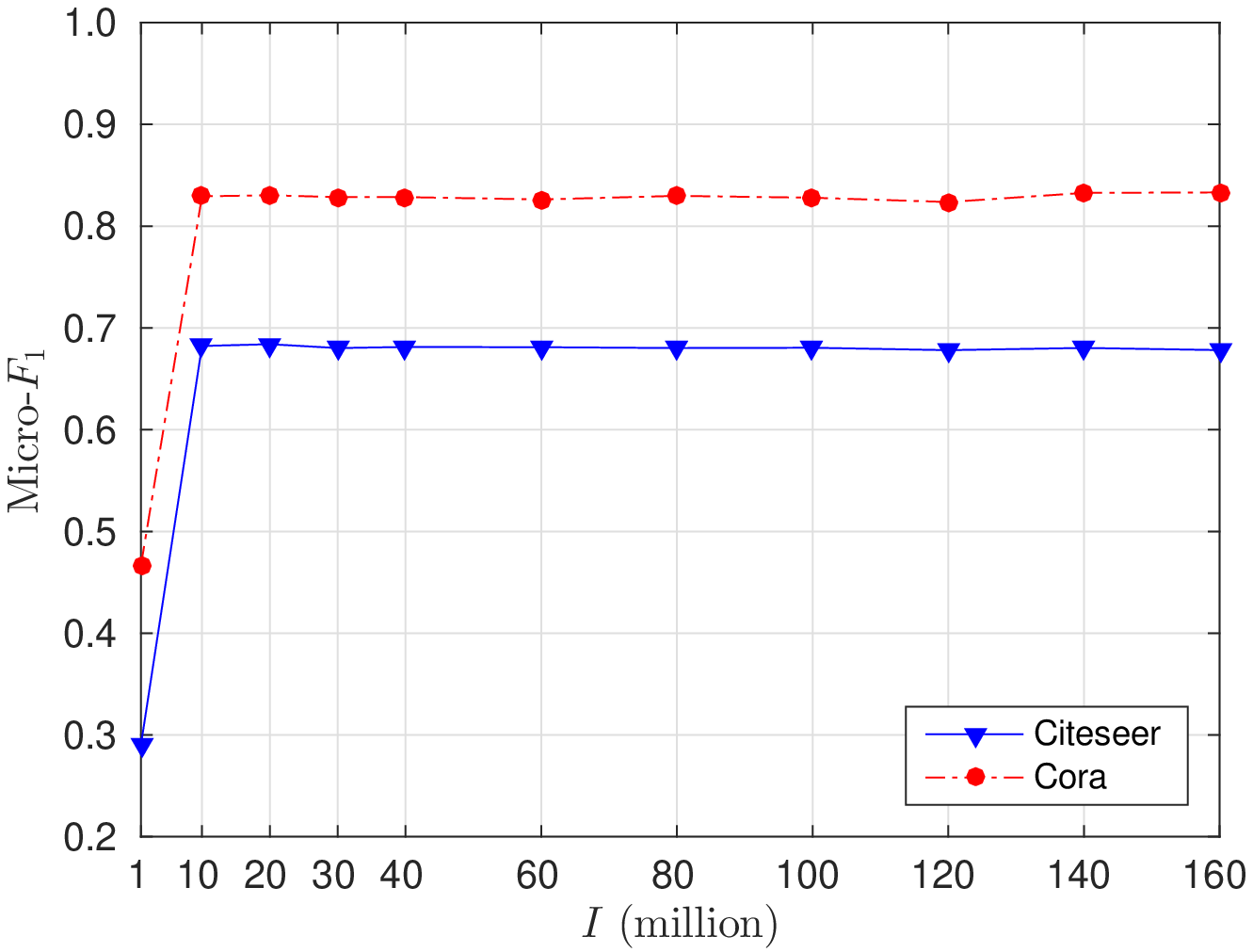}}
	\subfigure[$t$]{
		\label{fig:para:subfig:window}
		\includegraphics[width=2in]{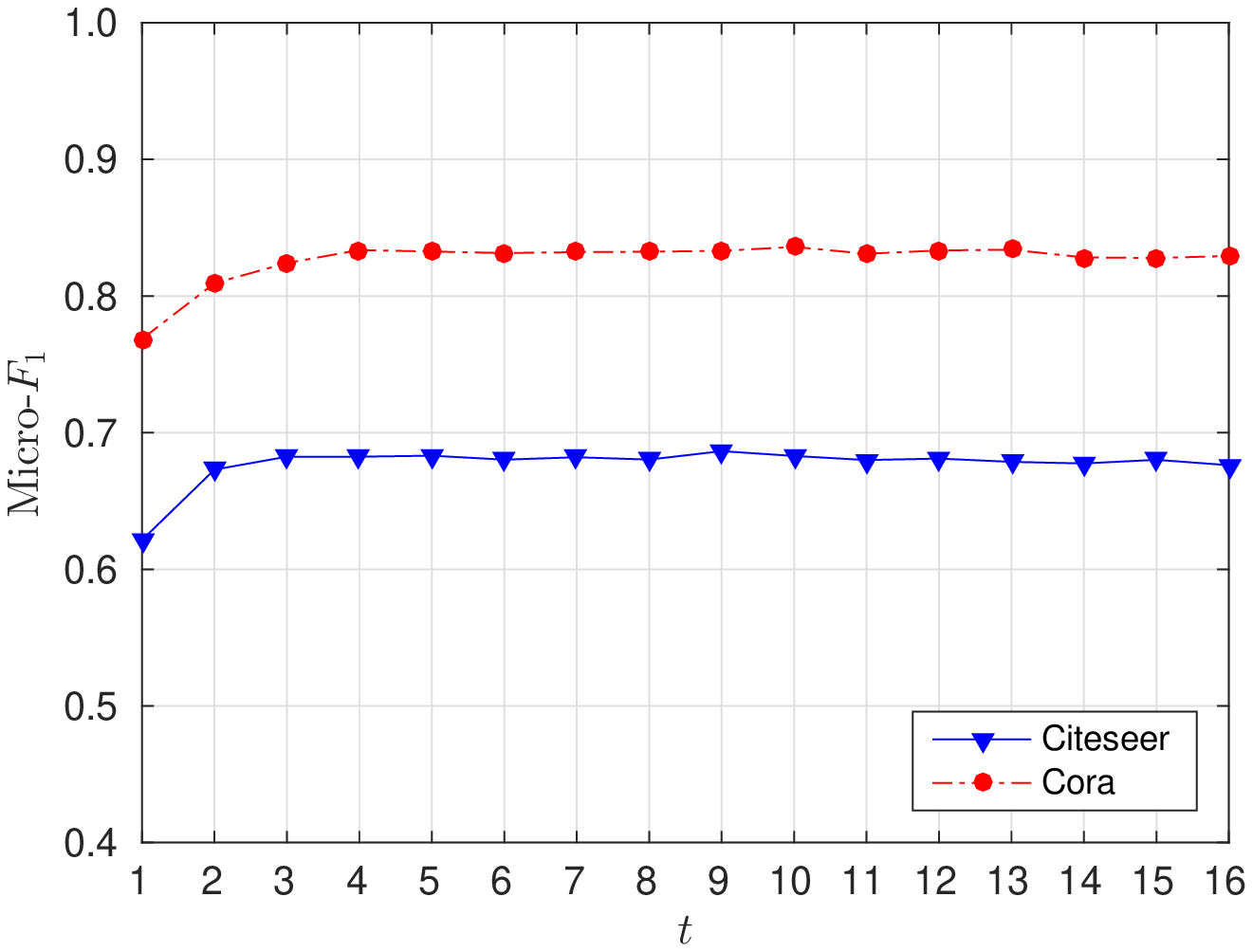}}
	\subfigure[$d$]{
		\label{fig:para:subfig:dim}
		\includegraphics[width=2in]{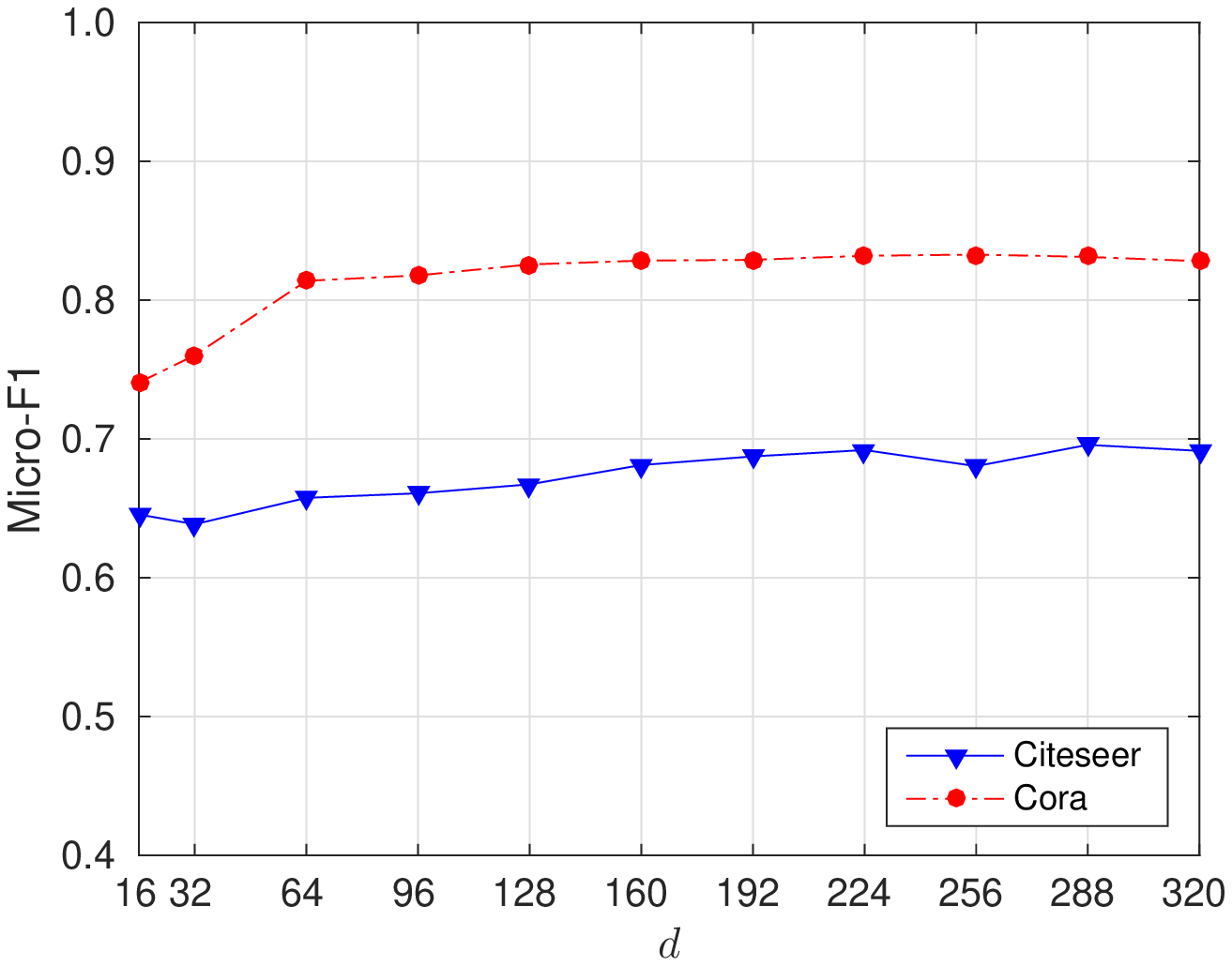}}
	\caption{Parameter sensitivity}
	\label{fig:para} 
\end{figure*}

Tables~\ref{LinkPrediction_Res_Cora}-\ref{LinkPrediction_Res_DBLPFull} report the link prediction results on Cora and DBLP(Full), with the best performer and the second best performer highlighted by \textbf{bold} and \underline{underline} respectively. As can be seen, on both Cora and DBLP(Full), the proposed SINE algorithm with the Hadamard operator consistently achieves the best link prediction performance with all missing edge ratios. We can also observe that, compared with attributed network embedding algorithms, the only structure preserving network embedding algorithms are more vulnerable to missing edges. As the ratio of missing edges increases, apart from the Attribute baseline, all methods experience a performance drop, while SINE retains more robust results. There are two main reasons: first, SINE uses random walks to bridge nodes with no immediate links and preserve their similarity in the learned node representations; second, SINE makes the best of the available node features to complement network structure for more effective network embedding.

\subsection{Experiments on Parameter Sensitivity}

We also conduct experiments to investigate the sensitivity of SINE to three important parameters: the maximum number of iterations $I$, window size $t$, and the dimension of learned node embeddings $d$.  We fix any two of the three parameters and study how the performance of SINE changes by varying the remaining one in turn. Fig.~\ref{fig:para} reports the change of Micro-$F_{1}$ values for node classification on Cora and Citeseer with regards to the three parameters, under the settings where attributes in randomly selected 50\% nodes are missing (random X row missing). Here, we observe a similar trend with an increase of all three parameters: SINE performs increasingly better and stabilizes after a threshold.

\subsection{Running Time Comparison}
\label{subsection: runningtime}

\begin{figure}[t]
	\centering
	\includegraphics[width=0.5\textwidth]{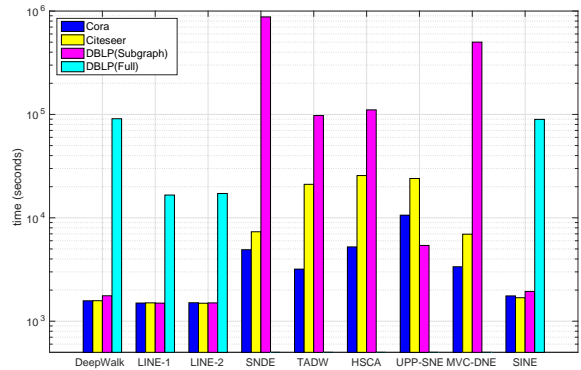}
	\caption{The comparison of running time of different network embedding algorithms on the log scale}
	\label{fig:time}
\end{figure}

We evaluate the running time of different network embedding algorithms in this section. We hope this empirical evaluation can help compare the efficiency and scalability of these algorithms, though they are implemented in different programming languages, with DeepWalk, UPP-SNE and SINE in C, LINE in C++, SDNE and MVC-DNE in Python, as well as TADW and HSCA in Matlab. Fig.~\ref{fig:time} compares the CPU running time (log-scale) of different  algorithms on Cora, Citeseer, DBLP(Subgraph) and DBLP(Full). On DBLP(Full), due to its large size, only the running time of DeepWalk, LINE-1, LINE-2 and SINE is available. We can see that, SINE is much more efficient than SDNE and other attributed network embedding methods including TADW, HSCA, UPP-SNE and MVC-DNE. The running time of SINE is comparable to that of DeepWalk, LINE-1 and LINE-2. This demonstrates high efficiency of SINE and its ability to scale to large-scale networks on a single machine (without GPU support).

\section{Conclusion}
Large-scale networks often contain incomplete node attributes and/or missing links, which impose significant challenges to attributed network embedding. Because missing node attributes or links result in inaccurate node similarity estimation, most existing methods are vulnerable to missing data, and minor presence of missing information may significantly deteriorate the algorithm performance. In this paper, we propose a novel scalable incomplete network embedding (SINE) method, which uses probabilistic learning of node-context and node-attribute relationships to tackle missing data on large-scale networks. SINE couples network node content and topology structures through a three-layer neural network, where each node learns its representation by considering context nodes and observable attributes of the node. By doing so, SINE fully minimizes the impact of missing data on the learning of node representations. A stochastic gradient descent based online algorithm is derived to ensure SINE can scale to large-scale networks. Extensive experiments and comparisons demonstrate the effectiveness and scalability of SINE for learning network embeddings on large-scale incomplete networks.
\label{sec-conclusion}

\section*{Acknowledgments}
The work was supported by the US National Science Foundation (NSF) through grant IIS-1763452, and the Australian Research Council (ARC) through grant LP160100630 and DP180100966. Daokun Zhang was supported by China Scholarship Council (CSC) with No. 201506300082 and a post-graduate scholarship from Data61, CSIRO in Australia.


%
%



%
%
%

\bibliographystyle{IEEEtran}
\bibliography{ICDM}

\begin{thebibliography}{10}
\providecommand{\url}[1]{#1}
\csname url@samestyle\endcsname
\providecommand{\newblock}{\relax}
\providecommand{\bibinfo}[2]{#2}
\providecommand{\BIBentrySTDinterwordspacing}{\spaceskip=0pt\relax}
\providecommand{\BIBentryALTinterwordstretchfactor}{4}
\providecommand{\BIBentryALTinterwordspacing}{\spaceskip=\fontdimen2\font plus
\BIBentryALTinterwordstretchfactor\fontdimen3\font minus
  \fontdimen4\font\relax}
\providecommand{\BIBforeignlanguage}[2]{{%
\expandafter\ifx\csname l@#1\endcsname\relax
\typeout{** WARNING: IEEEtran.bst: No hyphenation pattern has been}%
\typeout{** loaded for the language `#1'. Using the pattern for}%
\typeout{** the default language instead.}%
\else
\language=\csname l@#1\endcsname
\fi
#2}}
\providecommand{\BIBdecl}{\relax}
\BIBdecl

\bibitem{perozzi2014deepwalk}
B.~Perozzi, R.~Al-Rfou, and S.~Skiena, ``{DeepWalk}: Online learning of social
  representations,'' in \emph{Proc. of the 20th ACM SIGKDD Conf.}\hskip 1em
  plus 0.5em minus 0.4em\relax ACM, 2014, pp. 701--710.

\bibitem{tang2015line}
J.~Tang, M.~Qu, M.~Wang, M.~Zhang, J.~Yan, and Q.~Mei, ``{LINE}: Large-scale
  information network embedding,'' in \emph{Proc. of the 24th WWW Conf.}\hskip
  1em plus 0.5em minus 0.4em\relax ACM, 2015, pp. 1067--1077.

\bibitem{grover2016node2vec}
A.~Grover and J.~Leskovec, ``node2vec: Scalable feature learning for
  networks,'' in \emph{Proc. of the 22nd ACM SIGKDD Conf.}\hskip 1em plus 0.5em
  minus 0.4em\relax ACM, 2016, pp. 855--864.

\bibitem{yang2015network}
C.~Yang, Z.~Liu, D.~Zhao, M.~Sun, and E.~Y. Chang, ``Network representation
  learning with rich text information,'' in \emph{Proc. of the 24th IJCAI
  Conf.}, 2015, pp. 2111--2117.

\bibitem{zhang2016homophily}
D.~Zhang, J.~Yin, X.~Zhu, and C.~Zhang, ``Homophily, structure, and content
  augmented network representation learning,'' in \emph{Proc. of the 16th ICDM
  Conf.}\hskip 1em plus 0.5em minus 0.4em\relax IEEE, 2016, pp. 609--618.

\bibitem{yang2017properties}
D.~Yang, S.~Wang, C.~Li, X.~Zhang, and Z.~Li, ``From properties to links: Deep
  network embedding on incomplete graphs,'' in \emph{Proc. of the 26th ACM CIKM
  Conf.}\hskip 1em plus 0.5em minus 0.4em\relax ACM, 2017, pp. 367--376.

\bibitem{kossinets2006effects}
G.~Kossinets, ``Effects of missing data in social networks,'' \emph{Social
  Networks}, vol.~28, pp. 247--268, 2006.

\bibitem{zhang2018network}
D.~Zhang, J.~Yin, X.~Zhu, and C.~Zhang, ``Network representation learning: A
  survey,'' \emph{IEEE Transactions on Big Data}, 2018.

\bibitem{mikolov2013distributed}
T.~Mikolov, I.~Sutskever, K.~Chen, G.~S. Corrado, and J.~Dean, ``Distributed
  representations of words and phrases and their compositionality,'' in
  \emph{Prof. of the 27th NIPS Conf.}, 2013, pp. 3111--3119.

\bibitem{cao2015grarep}
S.~Cao, W.~Lu, and Q.~Xu, ``{GraRep}: Learning graph representations with
  global structural information,'' in \emph{Proc. of the 24th ACM CIKM
  Conf.}\hskip 1em plus 0.5em minus 0.4em\relax ACM, 2015, pp. 891--900.

\bibitem{wang2017community}
X.~Wang, P.~Cui, J.~Wang, J.~Pei, W.~Zhu, and S.~Yang, ``Community preserving
  network embedding.'' in \emph{Proc. of the 31th AAAI Conf. on Artificial
  Intelligence}, 2017, pp. 203--209.

\bibitem{cao2016deep}
S.~Cao, W.~Lu, and Q.~Xu, ``Deep neural networks for learning graph
  representations,'' in \emph{Proc. of the 30th AAAI Conf. on Artificial
  Intelligence}.\hskip 1em plus 0.5em minus 0.4em\relax AAAI Press, 2016, pp.
  1145--1152.

\bibitem{vincent2010stacked}
P.~Vincent, H.~Larochelle, I.~Lajoie, Y.~Bengio, and P.-A. Manzagol, ``Stacked
  denoising autoencoders: Learning useful representations in a deep network
  with a local denoising criterion,'' \emph{Journal of Machine Learning
  Research}, vol.~11, no. Dec, pp. 3371--3408, 2010.

\bibitem{wang2016structural}
D.~Wang, P.~Cui, and W.~Zhu, ``Structural deep network embedding,'' in
  \emph{Proc. of the 22nd ACM SIGKDD Conf.}\hskip 1em plus 0.5em minus
  0.4em\relax ACM, 2016, pp. 1225--1234.

\bibitem{natarajan2014inductive}
N.~Natarajan and I.~S. Dhillon, ``Inductive matrix completion for predicting
  gene--disease associations,'' \emph{Bioinformatics}, vol.~30, no.~12, pp.
  i60--i68, 2014.

\bibitem{wang2016paired}
S.~Wang, J.~Tang, F.~Morstatter, and H.~Liu, ``Paired restricted {Boltzmann}
  machine for linked data,'' in \emph{Proc. of the 25th ACM CIKM Conf.}\hskip
  1em plus 0.5em minus 0.4em\relax ACM, 2016, pp. 1753--1762.

\bibitem{hinton2006reducing}
G.~E. Hinton and R.~R. Salakhutdinov, ``Reducing the dimensionality of data
  with neural networks,'' \emph{Science}, vol. 313, no. 5786, pp. 504--507,
  2006.

\bibitem{zhang2017user}
D.~Zhang, J.~Yin, X.~Zhu, and C.~Zhang, ``User profile preserving social
  network embedding,'' in \emph{Proc. of the 26th IJCAI Conf.}, 2017, pp.
  3378--3384.

\bibitem{tu2017cane}
C.~Tu, H.~Liu, Z.~Liu, and M.~Sun, ``{CANE}: Context-aware network embedding
  for relation modeling,'' in \emph{Proc. of the 55th ACL Conf.}, vol.~1, 2017,
  pp. 1722--1731.

\bibitem{hamilton2017inductive}
W.~Hamilton, Z.~Ying, and J.~Leskovec, ``Inductive representation learning on
  large graphs,'' in \emph{Prof. of the 31st NIPS Conf.}, 2017, pp. 1024--1034.

\bibitem{huang2017accelerated}
X.~Huang, J.~Li, and X.~Hu, ``Accelerated attributed network embedding,'' in
  \emph{Prof. SDM}.\hskip 1em plus 0.5em minus 0.4em\relax SIAM, 2017, pp.
  633--641.

\bibitem{kim2011network}
M.~Kim and J.~Leskovec, ``The network completion problem: Inferring missing
  nodes and edges in networks,'' in \emph{Proc. of SDM}, 2011, pp. 47--58.

\bibitem{sundareisan2015hidden}
S.~Sundareisan, J.~Vreeken, and B.~A. Prakash, ``Hidden hazards: Finding
  missing nodes in large graph epidemics,'' in \emph{Proc. of SDM}, 2015, pp.
  415--423.

\bibitem{jr2009identifying}
G.~M.~S. Namata~Jr. and L.~Getoor, ``Identifying graphs from noisy and
  incomplete data,'' \emph{SIGKDD Explorations}, vol.~12, no.~1, pp. 33--39,
  2009.

\bibitem{smith2018network}
J.~M. Smith, Jeffrey~A. and J.~Morgan, ``Network sampling coverage {II}: The
  effect of non-random missing data on network measurement,'' \emph{Social
  Networks}, vol.~48, pp. 78--99, 2018.

\bibitem{li2014reducing}
A.~Q. Li, A.~Ahmed, S.~Ravi, and A.~J. Smola, ``Reducing the sampling
  complexity of topic models,'' in \emph{Proc. of the 20th ACM SIGKDD
  Conf.}\hskip 1em plus 0.5em minus 0.4em\relax ACM, 2014, pp. 891--900.

\bibitem{fan2008liblinear}
R.-E. Fan, K.-W. Chang, C.-J. Hsieh, X.-R. Wang, and C.-J. Lin, ``{LIBLINEAR}:
  A library for large linear classification,'' \emph{Journal of Machine
  Learning Research}, vol.~9, no. Aug, pp. 1871--1874, 2008.

\bibitem{strehl2002cluster}
A.~Strehl and J.~Ghosh, ``Cluster ensembles---a knowledge reuse framework for
  combining multiple partitions,'' \emph{Journal of machine learning research},
  vol.~3, no. Dec, pp. 583--617, 2002.

\end{thebibliography}

\end{document}